\documentclass{aastex63}

\usepackage{bm}

\newcommand\degree{^\circ}

\submitjournal{ApJ}

\shorttitle{}
\shortauthors{Luo et al.}

\graphicspath{{./}{figures/}}

\begin{document}
\title{Coherence of ion cyclotron resonance for damping ion cyclotron waves in space plasmas}

\correspondingauthor{Jiansen He}
\email{jshept@pku.edu.cn}

\author{Qiaowen Luo}
\affiliation{School of Atmospheric Sciences, Sun Yat-sen University, Zhuhai, 519000, China; cuijun7@mail.sysu.edu.cn; laihr@mail.sysu.edu.cn}
\affiliation{School of Earth and Space Sciences, Peking University, Beijing 100871, Beijing, China; jshept@pku.edu.cn}

\author{Xingyu Zhu}
\affiliation{School of Earth and Space Sciences, Peking University, Beijing 100871, Beijing, China; jshept@pku.edu.cn}

\author{Jiansen He}
\affiliation{School of Earth and Space Sciences, Peking University, Beijing 100871, Beijing, China; jshept@pku.edu.cn}


\author{Jun Cui}
\affiliation{School of Atmospheric Sciences, Sun Yat-sen University, Zhuhai, 519000, China; cuijun7@mail.sysu.edu.cn; laihr@mail.sysu.edu.cn}

\author{Hairong Lai}
\affiliation{School of Atmospheric Sciences, Sun Yat-sen University, Zhuhai, 519000, China; cuijun7@mail.sysu.edu.cn; laihr@mail.sysu.edu.cn}


\author{Daniel Verscharen}
\affiliation{Mullard Space Science Laboratory, University College London, Dorking RH5 6NT, UK}
\affiliation{Space Science Center, University of New Hampshire, Durham NH, USA}

\author{Die Duan}
\affiliation{School of Earth and Space Sciences, Peking University, Beijing 100871, Beijing, China; jshept@pku.edu.cn}

\begin{abstract}
Ion cyclotron resonance is one of the fundamental energy conversion processes through field-particle interaction in collisionless plasmas. However, the key evidence for ion cyclotron resonance (i.e., the coherence between electromagnetic fields and the ion phase space density) and the resulting damping of ion cyclotron waves (ICWs) has not yet been directly observed. Investigating the high-quality measurements of space plasmas by the Magnetospheric Multiscale (MMS) satellites, we find that both the wave electromagnetic field vectors and the bulk velocity of the disturbed ion velocity distribution rotate around the background magnetic field. Moreover, we find that the absolute gyro-phase angle difference between the center of the fluctuations in the ion velocity distribution functions and the wave electric field vectors falls in the range of (0, 90) degrees, consistent with the ongoing energy conversion from wave-fields to particles. By invoking plasma kinetic theory, we demonstrate that the field-particle correlation for the damping ion cyclotron waves in our theoretical model matches well with our observations. Furthermore, the wave electric field vectors ($\delta \mathbf{E'}_{\mathrm {wave,\perp}}$), the ion current density ($\delta \mathbf{J}_\mathrm {i,\perp}$) and the energy transfer rate ($\delta \mathbf{J}_\mathrm {i,\perp}\cdot \delta \mathbf{E'}_{\mathrm {wave,\perp}}$) exhibit quasi-periodic oscillations, and the integrated work done by the electromagnetic field on the ions are positive, indicates that ions are mainly energized by the perpendicular component of the electric field via cyclotron resonance. Therefore, our combined analysis of MMS observations and kinetic theory provides direct, thorough, and comprehensive evidence for ICW damping in space plasmas. 

\end{abstract}

\section{Introduction} \label{sec:intro}

Ion cyclotron waves are a prevalent phenomenon in various plasma environments, e.g., the Earth’s magnetosphere, the magnetosheath, and the solar wind \citep{Anderson1992, Dunlop2002, Usanova2012, Jian2009, He2011, Wicks2016, Zhao2018, Woodham2019, Telloni2019, Zhao2019, Bowen2020}. ICWs near the ion cyclotron frequency can have a close coupling with ions through cyclotron resonance. ICWs are regarded as one of the crucial wave modes in shaping the particle kinetics locally (plasma ions and energetic electrons) and even the dynamics of the global magnetospheric system \citep{Thorne2010, Yuan2014, Su2014}. ICWs can have different wavebands corresponding to the cyclotron frequencies of different ion species (e.g., $H^+$, $He^+$, $O^+$), and, in the magnetospheric context, may be located in different regions in terms of L-shell and magnetic local time (MLT) \citep{Allen2015, Wang2017}. In the Earth’s magnetosphere, ICW can be generated by temperature-anisotropy instabilities through releasing the excess of the ion perpendicular thermal energy, in which case the wave amplitude saturates when the ion thermal anisotropy approaches an equilibrium state. It is widely believed that ICWs cause precipitation of relativistic electrons and energetic ions from the magnetosphere down to the ionosphere and atmosphere through pitch angle scattering \citep{Zhang2016, Hendry2016, Kurita2018, Qin2018}, contributing to the decay phase of geomagnetic storms \citep{Jordanova2006}. ICWs can also be damped by converting energy from waves to particles. For example, ICWs can accelerate ions through cyclotron resonance in the polar region, leading to the loss of $O^+$ from the Earth’s atmosphere \citep{Chang1986},  or heat thermal ions preferentially in the direction perpendicular to the background field \citep{Marsch2006}. Quantification of the wave-particle interactions and the association of energy transfer between waves and particles is necessary to better understand critical space plasma phenomena such as ion kinetic physics, particle precipitation, the atmospheric loss processes, and the evolution of geomagnetic storms.

Identifying the resonance mechanisms that convert energy between electromagnetic fields and charged particles in nearly collisionless plasmas is a critical step to understand the process of wave–particle interactions \citep{Hollweg2002, He2015, Verscharen2019}.  \citet{He2015} revealed the coexistence of two wave modes (quasi-parallel ICWs and quasi-perpendicular kinetic Alfvén waves (KAWs)) and three resonance diffusion plateaus in proton velocity space, which suggests a complicated scenario of wave–particle interactions in solar wind turbulence: left-handed cyclotron resonance between ICWs and the proton core population, and Landau and right-handed cyclotron resonances between KAWs and the proton beam population. According to kinetic theory, ions can be energized by the perpendicular component of the electric field in a sub-region of velocity space via cyclotron resonance \citep{Duan2020, Klein2020}. For the energy transfer via Landau resonance, the field-particle correlation method has been successfully implemented to explore compressive waves in simulations \citep{Ruan2016, Klein2016, Howes2018} as well as in observations \citep{Chen2019}.  As a pioneering effort of seeking observational evidence for cyclotron resonance, \citet{Kitamura2018} find that the observed ion differential energy flux spectra are not symmetric around the magnetic field direction but are in phase with the plasma wave fields, suggesting that the energy is transferred from ions to ion cyclotron waves via cyclotron resonance. 

The $\mathbf{J}\cdot \mathbf{E'}$ term is often studied in observational time series and in simulation data to quantify the energy transfer between fields and particles at various scales  \citep{Yang2017, Chasapis2018, He2019, He2020, Luo2020, Duan2020}. For the interaction between ions and waves, the energy transfer rate is calculated as the dot product of the fluctuating electric field ($\mathbf{E'}_{wave}$) and the fluctuating ion current ($\mathbf{J}_\mathrm i$), both of which are perpendicular to the background magnetic field $\mathbf{B}_0$ in cyclotron-resonant interactions \citep{Omura2010}. Aside from the $\mathbf{J}\cdot \mathbf{E'}$ term, the term for the pressure–strain tensor interaction, $-(\mathbf{P}\cdot \nabla)\cdot \mathbf{V}$, is another proxy for energy dissipation, representing the energy conversion from bulk kinetic energy to thermal energy \citep{Yang2017, Chasapis2018, Luo2020}. Simulations suggest that, although scale-dependent, the spatial patterns of $\mathbf{J}\cdot \mathbf{E'}$ and $-(\mathbf{P}\cdot \nabla)\cdot \mathbf{V}$ are often concentrated in proximity to each other \citep{Yang2019}. 

However, the field-particle coherent interaction, which is responsible for the damping of ion cyclotron waves has not been directly observed. More specifically, the details of the interaction between the electromagnetic field of the ion cyclotron waves with the fluctuating ion velocity distribution function $(\delta f_{\mathrm i}(\mathbf V))$ is of great importance for understanding field-particle interactions. Here, we present the first observation of the correlation between the ion velocity distribution function and the wave electric field vectors elucidating the process of field-particle interaction and the damping process of ion cyclotron waves.

\section{Observations of ICWs and associated field-particle correlation}\label{sec:data}

We survey the ICW list from the website of the MMS science data center\footnote{https://lasp.colorado.edu/mms/sdc/public/}, and select the events that were observed in the magnetosphere in burst mode. As a result, we acquire 44 ICW events from October 15, 2018 to March 13, 2021. We find ICW growth in 17 of the 44 events, and ICW damping in 16 of the 44 events. The remaining 11 events have no clear signal of ICW growth or damping. Since the growth of these listed ICW has been studied in depth before \citep[e.g.,][]{Kitamura2018}, here we focus on the damping of ICWs. The event, which has typical and clear coherent coupling features between fields and particles, is selected in this study for detailed analysis. It was encountered on 2018 November 01 at 16:39:03 UT-16:39:52 UT, when the MMS spacecraft \citep{Burch2016} were in the Earth’s outer magnetosphere, and near the magnetopause (Figure 1(a)). We use data from the magnetometer at 128 samples/s (Fluxgate Magnetometer) \citep{Russell2016}, the Fast Plasma Investigation (FPI) \citep{Pollock2016} at 150ms for ions. By employing the singular value decomposition (SVD) of the electromagnetic spectral matrix according to Gauss’s and Faraday’s laws \citep{Santolk2003}, we find the fluctuations are left-hand circularly polarized about the local mean magnetic field direction ($\mathbf{B}_{\mathrm{0,local}}$) and propagate anti-parallel to $\mathbf{B}_{\mathrm{0,local}}$ (Figure 1(b)-(c)), strongly suggesting their nature as ion cyclotron waves (ICWs). In this ICW event, since the period of ICWs is $\sim$4s, only the time resolution of burst intervals in magnetic field measurements (128 samples/s) and ions measurements (150ms) can meet the needs of analysis. Since it lacks the burst mode data of magnetic field and particles before and after the interval between 16:39:03 UT and 16:39:52 UT, we choose the time period from 16:39:03 UT to 16:39:52 UT on November 01, 2018 for further analysis.

\begin{figure}
\centerline{\includegraphics[width=13cm, clip=]{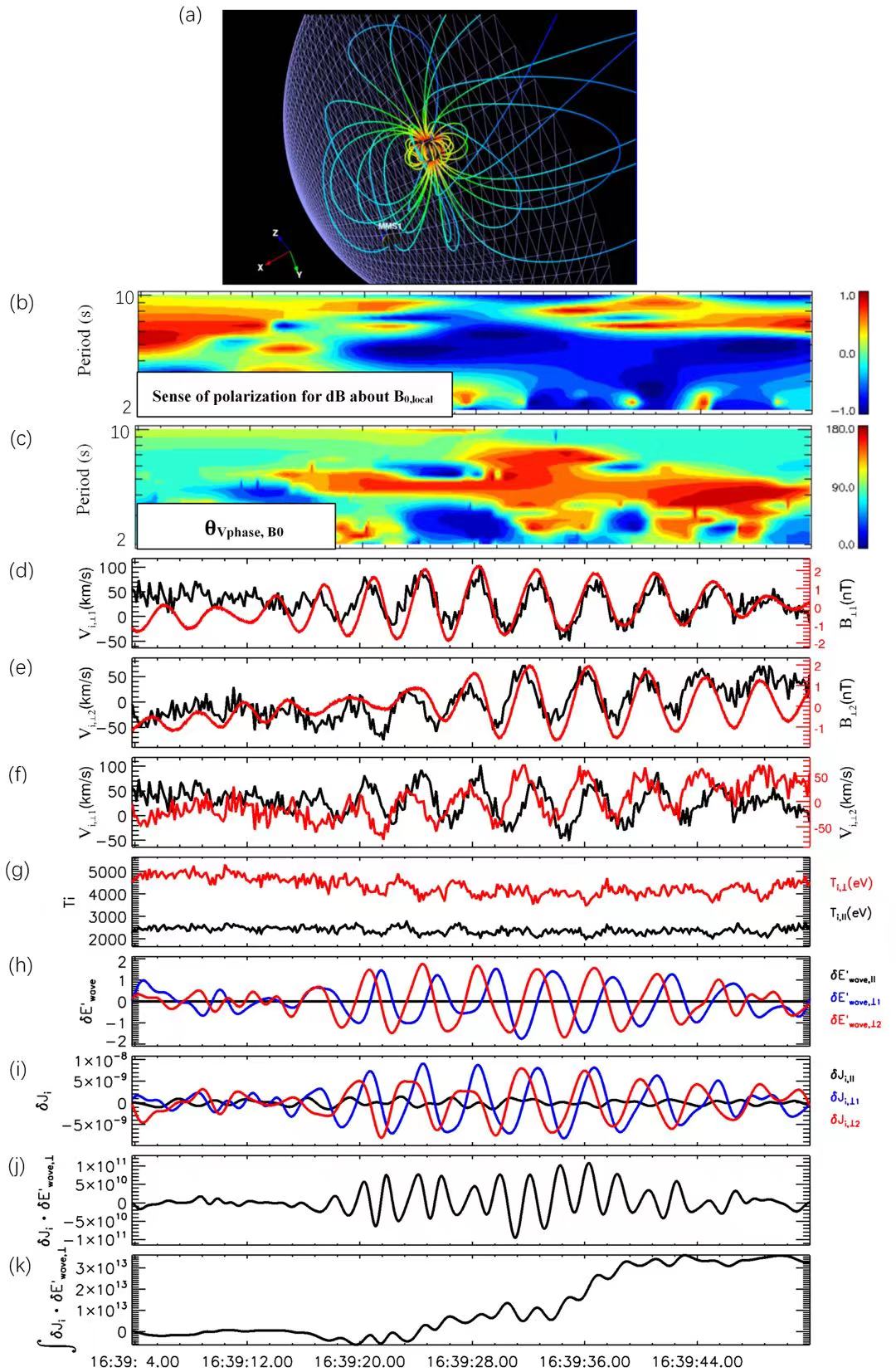}}
\caption{Alfvénic cyclotron waves in the Earth's outer magnetosphere near the magnetopause as measured by MMS on 2018 November 01. (a) Position of MMS in the outer magnetosphere during the measurement. It is located inside the magnetopause. (b) Sense of polarization for $\delta \mathbf{B}_{\perp}$ around the local mean magnetic field direction $\mathbf{B}_{0,local}$, with the values of $-1$ and $+1$ representing the left-hand and right-hand circular polarization about $\mathbf{B}_{0,local}$. (c) Angle of propagation direction for electromagnetic field fluctuations with respect to $\mathbf{B}_{0,local}$, $\theta_{Vphase,B0}$. (d-f) Time series of magnetic field components ($\mathbf{B}_{\perp1}$ and $\mathbf{B}_{\perp2}$) as well as ion (proton) bulk/fluid velocity components ($\mathbf{V}_{\mathrm i,\perp1}$ and $\mathbf{V}_{\mathrm i,\perp2}$) in field-aligned coordinates. The positive correlation between the $\mathbf{B}$-component and the $\mathbf{V}$-component indicates an anti-parallel propagation of ion cyclotron waves. The phase of $\mathbf{V}_{\mathrm i,\perp1}$ is 90$\degree$ ahead of $\mathbf{V}_{\mathrm i,\perp2}$, indicates the left-hand polarization. (g) The parallel and perpendicular temperatures of protons. (h$\&$i) The wave electric field components and wave ion current density components. (j) The work done by the wave electric fields on the ions in the perpendicular direction. (k) The time-integrated work done by the wave electric field on the ions in the perpendicular direction. 
\label{fig:fig1}}
\end{figure}

We use the background magnetic field ($\mathbf{B}_{\mathrm 0}$, i.e., the magnetic field averaged over the full-time interval) to define the magnetic field–aligned coordinates. The subscript $\perp2$ denotes the $\mathbf{B}_{\mathrm 0}\times\mathbf{V}_{\mathrm 0}$ direction (where $\mathbf{V}_{\mathrm 0}$ is the velocity averaged over the full-time interval), and the subscript $\perp1$ completes the right-handed system. In Figure 1(d)-(e), the positive correlation between the B-component $\mathbf{B}_{\mathrm \perp1}$($\mathbf{B}_{\mathrm \perp2}$) and the V-component $\mathbf{V}_{\mathrm i,\perp1}$($\mathbf{V}_{\mathrm i,\perp2}$) indicates an anti-parallel propagation of ICWs. In Figure 1(f), the phase of $\mathbf{V}_{\mathrm i,\perp2}$ is $90\degree$ ahead of $\mathbf{V}_{\mathrm i,\perp1}$, indicates the left-hand polarization of the wave mode. The temperature of the ions (Figure 1(g)) shows a thermal anisotropy, with $T_{\mathrm i,\perp} /T_{\mathrm i,\parallel}=1.83$.

In Figure 2, the power spectral densities (PSDs) of ion (proton) bulk/fluid velocity (Figure 2(a)-(c)), the magnetic field (Figure 2(d)-(f)), and the electric field (Figure 2(g)-(i)), as well as the spectra of the energy transfer rate (Figure 2(j)-(l)), show peaks around 0.25Hz. Hence, we define the wave magnetic field ($\delta \mathbf B_{\mathrm wave}$) as the magnetic field in the frequency range of 0.1–0.5Hz, and obtain the wave magnetic field through filtering with the inverse Fast Fourier transform. The components of $\mathbf{V}_{\mathrm i}$ in the frequency range of 0.1–0.5 Hz ($\delta \mathbf{V}_{\mathrm{i,wave}}$) are filtered in the same way as $\delta \mathbf{B}_{\mathrm wave}$. The filtered wave electric field vectors ($\delta \mathbf{E'}_{\mathrm wave}$) in the plasma frame, which moves with the mean flow velocity, is plotted in Figure 1(h) and the phase of $\delta \mathbf{E'}_{\mathrm wave,\perp2}$ is $90\degree$ ahead of $\delta \mathbf{E}'_{\mathrm wave,\perp1}$. In Figure 1, the time series of the magnetic field and the fluid velocity are the original measurements, while the wave electric field components and wave ion current density components are filtered in the frequency range of 0.1–0.5~Hz.

\begin{figure}
\centerline{\includegraphics[width=13cm, clip=]{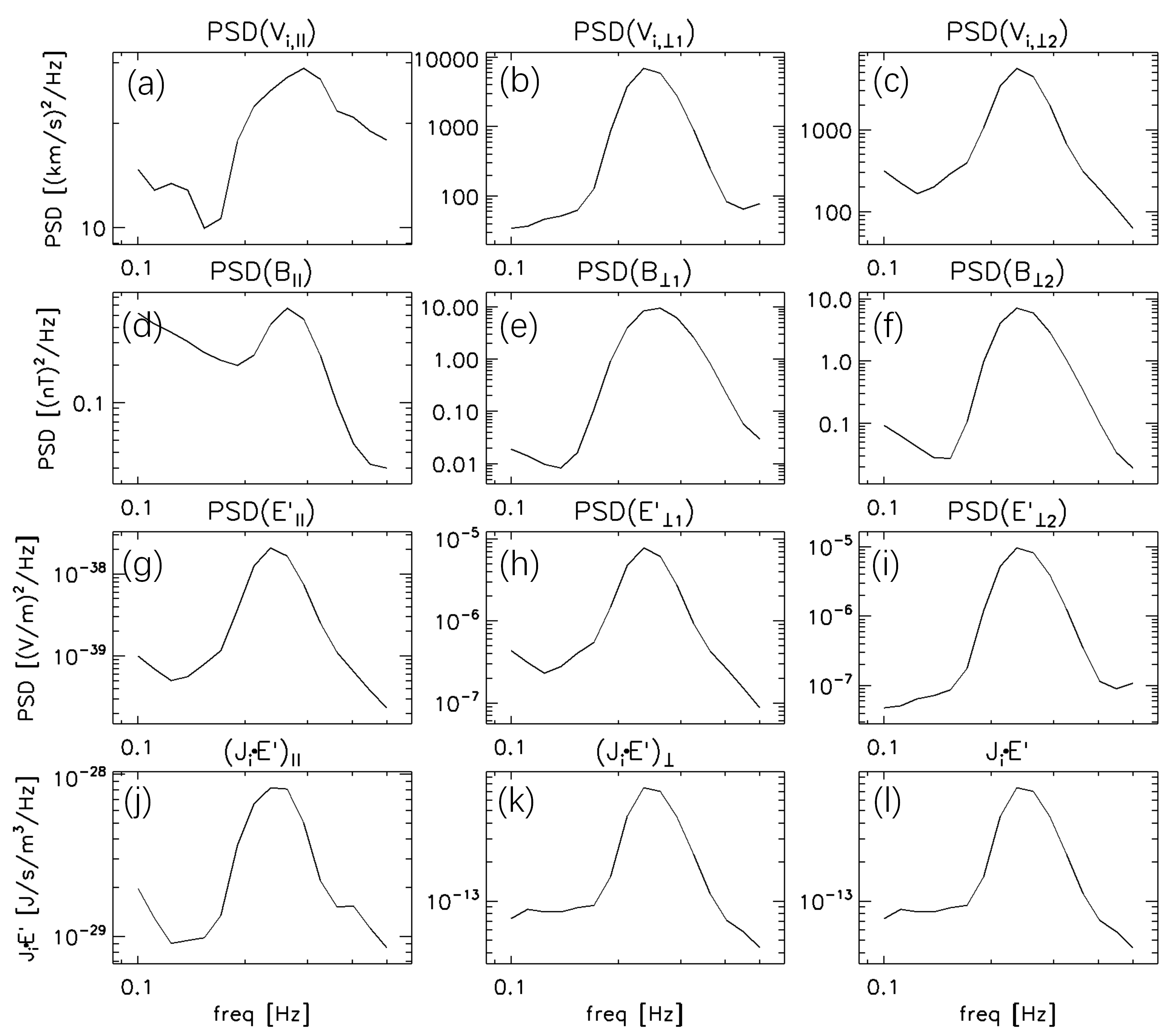}}
\caption{(a-c) The power spectral density (PSD) of  ion (proton) bulk/fluid velocity components ($\mathbf{V}_{\mathrm i,\parallel}$, $\mathbf{V}_{\mathrm i,\perp1}$ and $\mathbf{V}_{\mathrm i,\perp2}$) in field-aligned coordinates.  (d-f) The power spectral density (PSD) of  magnetic field components ($\mathbf{B}_{\parallel}$ , $\mathbf{B}_{\perp1}$ and $\mathbf{B}_{\perp2}$ ) in field-aligned coordinates. (g-i) The power spectral density (PSD) of  electric field components ($\mathbf{E}_{\parallel}$ , $\mathbf{E}_{\perp1}$ and $\mathbf{E}_{\perp2}$ ) in field-aligned coordinates. (j-l) Spectrum of the energy transfer rate ($ ({\mathbf{J}_\mathrm {i}\cdot  \mathbf{E'}})_\parallel$, $ ({\mathbf{J}_\mathrm {i}\cdot  \mathbf{E'}})_{\perp}$ and ${\mathbf{J}_\mathrm {i}\cdot  \mathbf{E'}}$) in field-aligned coordinates.
\label{fig:fig2}}
\end{figure}

The energy transfer rate via cyclotron-resonant interactions between ICWs and ions is calculated as the dot product of $\delta \mathbf{E'}_{\mathrm wave}$ and the fluctuating ion current density ($\delta \mathbf{J}_{\mathrm i}$) perpendicular to $\mathbf{B}_{\mathrm 0}$. The contributions to the total current density from the ion species are calculated as $\mathbf{J}_{\mathrm i} = N_{\mathrm i}\cdot q_{\mathrm i}\cdot \mathbf{V}_{\mathrm i}$ ($N_{\mathrm i}$ is ion's number density, $q_{\mathrm i}$ is ion's charge, $\mathbf{V}_{\mathrm i}$ is ion's bulk velocity) and the filtered fluctuating ion current density ($\delta \mathbf{J}_{\mathrm i}$) is plotted in Figure 1(i). The work done by the electromagnetic field on the ions in the perpendicular directions is illustrated in Figure 1(j). Lastly, the integrated work done by the electromagnetic field on the ions is shown in Figure 1(k). We note that the wave electric field vectors ($\delta \mathbf{E'}_{\mathrm {wave,\perp}}$), the ion current density ($\delta \mathbf{J}_\mathrm {i,\perp}$) and the energy transfer rate ($\delta \mathbf{J}_\mathrm {i,\perp}\cdot \delta \mathbf{E'}_{\mathrm {wave,\perp}}$) exhibit quasi-periodic oscillations. Positive $\delta \mathbf{J}_{\mathrm {i,\perp}}\cdot \delta \mathbf{E'}_{\mathrm wave,\perp}$  indicates that ions are mainly energized by the perpendicular component of the electric field via cyclotron resonance. In this event, the trend of $B_z$ is similar to the trend of the time-integrated work done by the wave electric field on the ions in the perpendicular direction. Moreover, our interpretation of an active wave-particle interaction also requires certain phase relations between the electric field and the fluctuations in the particle distribution. If we reverse the time series and conduct the same analysis on the new time series, we find that the time-integrated work still has an increasing trend, while the trend of $B_z$ is decreasing, which furthermore suggests an active wave-particle interaction.

First-order left-hand cyclotron resonance occurs when the resonance condition $\omega - k_{\parallel}\mathbf{V}_{\parallel} = n\Omega_{\mathrm i}$ is satisfied with n=1, where $\omega$ is the wave frequency in the plasma frame, $k_{\parallel}$ is the wavenumber component parallel to $\mathbf{B}_0$, $\mathbf{V}_{\parallel}$ is the particle parallel velocity component, $\Omega_{\mathrm i}$ is the proton cyclotron frequency, and $n\neq0 $ is the integral resonance number. For the MMS observation considered here, because the ICWs propagate anti-parallel to $\mathbf{B}_{\mathrm 0}$ and the frequency of ICWs ($\omega\sim1.28rad/s$) is smaller than the proton gyro-frequency ($\Omega_{\mathrm i}=3.57rad/s$), the resonance condition is satisfied for ions with pitch angles smaller than $90\degree$.

The correlation between the azimuthal angle of the center of the fluctuating ion phase space density $\phi(\delta f_{\mathrm i}(t))$ and the azimuthal angle of the wave electric field vectors $\phi(\delta \mathbf{E'}_{\mathrm {wave,\perp}}(t))$ is shown in Figure 3(a)-(d). Figure 3(a)-(d) illustrate the $(t-\phi)$ diagrams of the fluctuating ion phase space density ($\delta f_{\mathrm i}$  = $f_{\mathrm i}$ - $<f_{\mathrm i}>$, where $<f_{\mathrm i}>$ is the time-average of the ion phase space densities) at energies from 2 eV to 17013 eV and at pitch angles from 2° to 87° as the background. We then superpose the time series of $\phi(\delta \mathbf{E'}_{\mathrm {wave,\perp}}(t))$ on them. In other words, from 16:39:18 to 16:39:46, the observed ion velocity distributions are not symmetric around the magnetic field direction but are in phase with the plasma wave fields. Moreover, the absolute angle difference between the azimuthal angle of the fluctuating ion phase space density $\phi(\delta f_{\mathrm i}(t))$, which can be approximated with the angle of the time-dependent local maximum $\phi^*$ satisfying $\delta f_{\mathrm i}(t,\phi^*)=\delta f_{\mathrm{i,max}}(t)$, and the azimuth angle of wave electric field vectors $\phi (\delta \mathbf{E'}_{\mathrm {wave,\perp}}(t))$ is less than $90\degree$. Such a positive correlation between $\phi(\delta \mathbf{E'}_{\mathrm {wave,\perp}}(t))$  and $\phi(\delta f_{\mathrm i}(t))$ is consistent with positive work done by the electromagnetic field in Figure 1 from 16:39:18 to 16:39:46.

\begin{figure}
\centerline{\includegraphics[width=13cm, clip=]{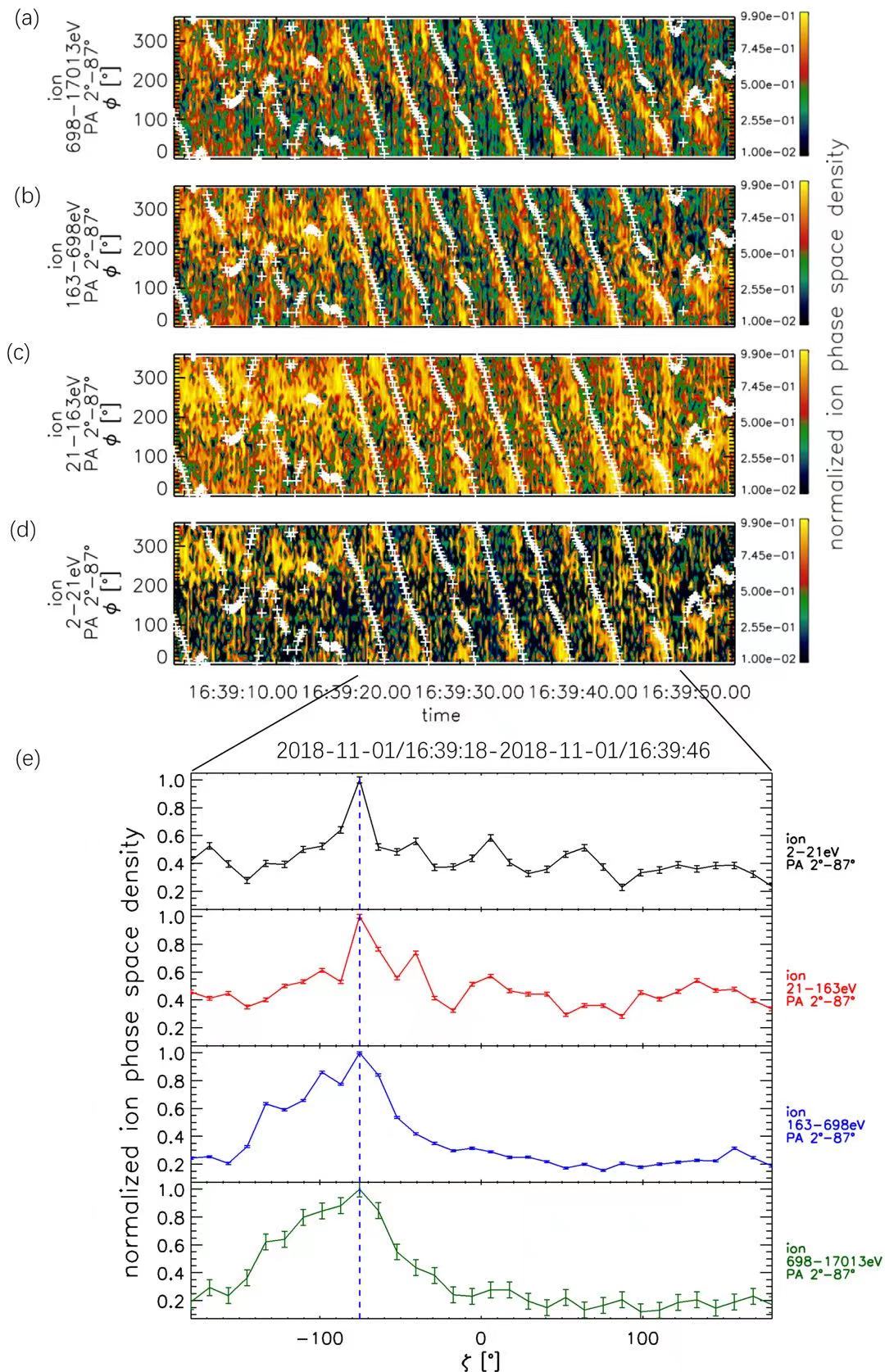}}
\caption{(a-d) Correlation between the azimuthal angle of the center of the fluctuating ion phase space density $\phi(\delta f_{\mathrm i}(t))$, which can be approximated with the angle of the time-dependent local maximum $\delta f_{\mathrm i}(t,\phi)$, and the azimuthal angle of the wave electric field vectors $\delta \mathbf{E'}_{\mathrm {wave,\perp}}$ (white crosses). The observed ion distributions are not symmetric around the magnetic field direction but are in phase with the plasma wave fields. (e) Relative phase angle ($\zeta$) (i.e., the gyro phase relative to the rotating $\delta \mathbf{E'}_{\mathrm {wave,\perp}}$) distributions of the 28s-averaged (from 16:39:18 to 16:39:46) ion phase space densities with the error bars ($\sigma$) representing twice the standard error of the mean at every angular bin.  $\sigma$= $\sqrt{\sum_{i=1}^n{(f_{\mathrm i}(t,\zeta)-<f(t,\zeta)>)^2}/n}$, where n=7 is the number of waves periods from 16:39:18 to 16:39:46. $f_{\mathrm i}(t,\zeta)$ is the average ion phase density in the $i^{th}$ period and $<f(t,\zeta)>$ is the average ion phase density of 7 periods.
The blue dashed vertical line in Figure 3(e) denotes the normalized ion phase space density peak around $\zeta=-75\degree$.
\label{fig:fig3}}
\end{figure}

To investigate the damping cyclotron resonance during the interval of [16:39:18, 16:39:46] in more detail, we sort the data of the ion phase space density according to the relative phase angle ($\zeta$), which is defined as the azimuthal angle difference between $\phi(\delta f_{\mathrm{i,max}})$ and $\phi(\delta \mathbf{E'}_{\mathrm {wave,\perp}})$ to represent the gyro phase difference $\delta f_{\mathrm{i,max}}$ relative to the rotating $\delta \mathbf{E'}_{\mathrm {wave,\perp}}$. The normalized ion phase space densities as functions of the relative phase angle ($\zeta$) averaged over the time duration of 28s are shown in Figure 3(e), where a significant peak around $\zeta$= -75° can be identified at all energies. Again the absolute relative phase angle $|\zeta|$ = $75\degree$ is less than $90\degree$, clearly suggesting an ongoing process of energy transfer from fields to particles and the damping of wave electromagnetic field energy.

\section{Comparison of field-particle correlation between observation and theory}\label{sec:data}
To compare with the observational results, we investigate the field-particle correlation of ion cyclotron waves based on linear plasma wave theory \citep{Stix1992} using our newly developed solver for the full set of perturbations of the linear plasma eigenmodes (Plasma Kinetics Unified Eigenmode Solutions, PKUES). The first part of PKUES is inherited from the solver “Plasma Dispersion Relation Kinetics” (PDRK) \citep{Xie2016} and calculates all possible eigenmode solutions at a time. Furthermore, like “NHDS” \citep{Verscharen2018}, PKUES provides a full set of characteristic fluctuations (including $\delta \mathbf B$, $\delta \mathbf E$, $\delta f_{\mathrm i}$, $\delta f_{\mathrm e}$, $\delta N_{\mathrm i}$, $\delta \mathbf V_{\mathrm i}$, and $\delta \mathbf  V_{\mathrm e}$) for the eigenmode under study. By applying the observed plasma conditions to PKUES, we calculate the coherent fluctuating phase space density of the specific mode as a function of time, and illustrate it in Figure 4 after adding the background bi-Maxwellian distribution. The magnetized plasma parameters used in PKUES are: $n_{\mathrm e0}=n_{\mathrm p0}=1.1 cm^3$, $T_{\mathrm pi}$=2383.3 eV, $T_{\mathrm ei}$=348.0 eV, $T_{\mathrm p\perp}$=4340.6 eV, $T_{\mathrm e\perp}$=458.3 eV, $\mathbf{v}_{\mathrm d,e}$=$\mathbf{v}_{\mathrm d,i}$=0 km/s. The theoretically-predicted azimuthal angle correlation between the fluctuating ion phase space density ($\delta f_{\mathrm i}(t)$) and the wave electric field vectors ($\delta \mathbf{E'}_{\mathrm {wave,\perp}}(t))$ also suggests a field-to-particle energy transfer as we observe in Figure 3. Likewise, the relative phase angle ($\zeta$) distributions of the theoretically-predicted ion phase space densities are shown in Figure 4e. We observe a peak around $\zeta=-75\degree$ located between $[-90, 0]\degree$ in Figure 4(e), demonstrating that a cyclotron resonance transfers energy from the waves to the ions. 

\begin{figure}
\centerline{\includegraphics[width=13cm, clip=]{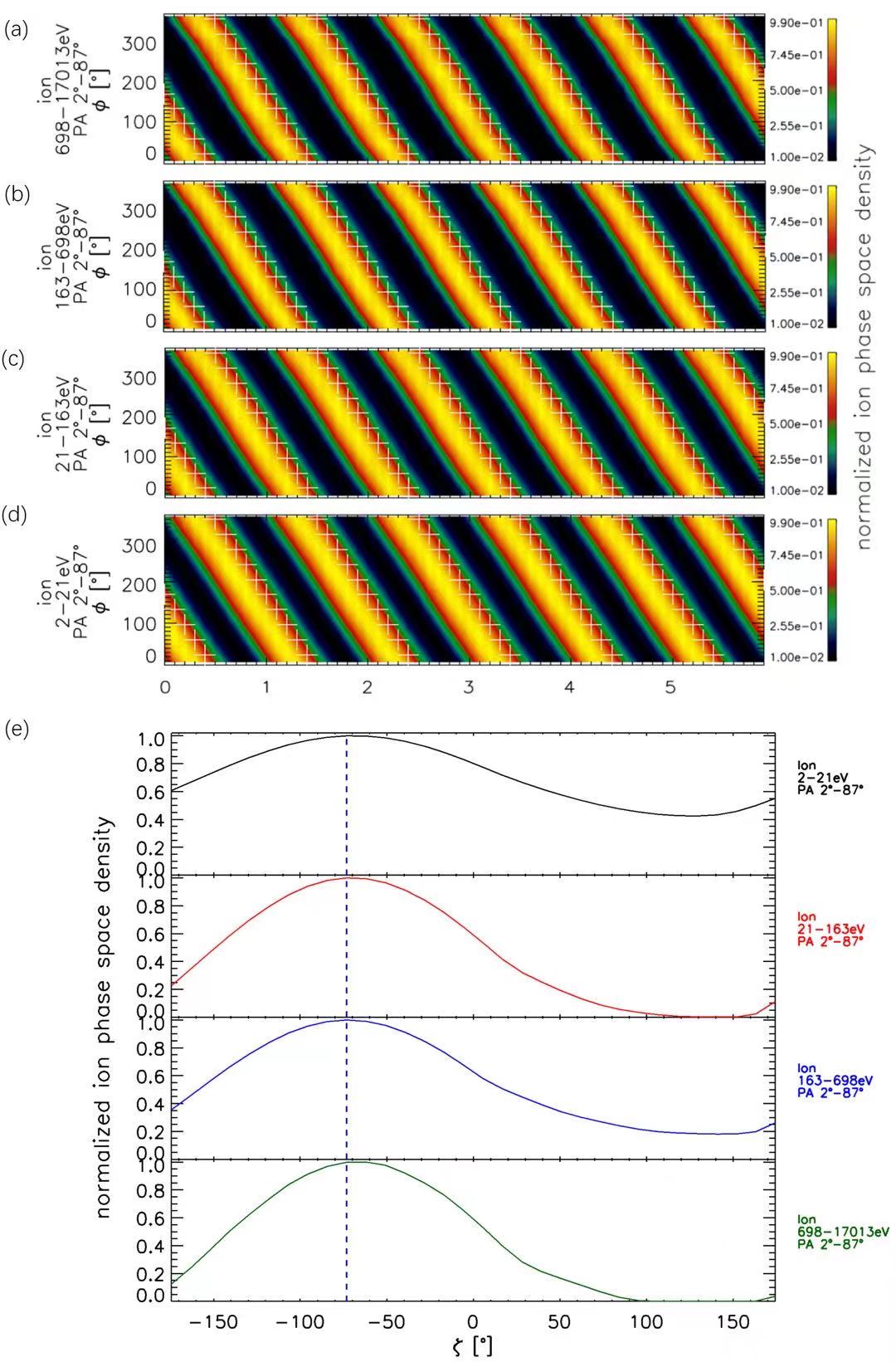}}
\caption{(a-d) Correlation between the azimuthal angle of the center of the fluctuating ion phase space density $\phi(\delta f_{\mathrm i}(t))$, the proxy to which can be the angle of the time-dependent local maximum of $\delta f_{\mathrm i}(t,\phi)$, and the azimuthal angle of the wave electric field vector $\delta \mathbf{E'}_{\mathrm {wave,\perp}}$, based on the plasma wave prediction from the PKUES solver. (e) Relative average phase angle ($\zeta$) distributions of the ion phase space densities. The blue dashed vertical line in Figure 4(e) denotes the normalized ion phase space density peak around $\zeta =-75\degree$. 
\label{fig:fig4}}
\end{figure}

In Figure 5 and Figure 6, we illustrate how the cyclotron wave’s electromagnetic field vectors ($\delta \mathbf{E'}_{\mathrm {wave,\perp}}$ and $\delta \mathbf{B}_{\mathrm {wave,\perp}}$) correlate with the fluctuating ion velocity distribution function ($\delta f_{\mathrm i}(\mathbf{V})$). Here, we focus on the energy transfer from ICWs to the ions, which are recorded at a time cadence of 150ms, resulting in a total of 327 snapshots of three-dimensional velocity distributions. The fluctuating ion velocity distribution function is calculated as $\delta f_{\mathrm i}(\mathbf V)$ = $f_{\mathrm i}(\mathbf V)$ - $<f_{\mathrm i}(\mathbf V)>$, where $<f_{\mathrm i}(\mathbf V)>$ is the average of these 327 three-dimensional velocity distributions. We show $\delta f_{\mathrm i}(\mathbf V)$ (the blue contour surface), $\delta \mathbf{B}_{\mathrm {wave,\perp}}$ (the yellow sticks) and $\delta \mathbf{E'}_{\mathrm {wave,\perp}}$ (the green sticks) during the period from 16:39:32.63 to 16:39:35.93 of the ICW event in Figure 5. The central position (i.e., the bulk velocity) of $\delta f_{\mathrm i}(\mathbf V)$ is mostly in phase with $\delta \mathbf{E'}_{\mathrm {wave,\perp}}$ and the angle between them is less than $90\degree$ for most of the times shown. The phase relations between the wave fields and ions demonstrate that the cyclotron resonance transfers energy from the wave fields to the ions.

In Figure 6, the 3-D contour surfaces of the fluctuating (opaque) and total (transparent) ion phase space densities at different phases in one wave period are shown based on the PKUES solver.  From Figure 6(a) to Figure 6(h),  the bulk velocity vector of  $\delta f_{\mathrm i}(\mathbf V)$ rotates with $\delta \mathbf{E'}_{\mathrm {wave,\perp}}$ in the sense of left-hand polarization. Moreover, the 3-D contour of $\delta f_{\mathrm i}(\mathbf V)$ bends towards the direction parallel to $\delta \mathbf{E'}_{\mathrm {wave,\perp}}$. Such a co-rotation of the agyrotropic $\delta f_{\mathrm i}(\mathbf V)$ with $\delta \mathbf{B}_{\mathrm {wave,\perp}}$ and $\delta \mathbf{E'}_{\mathrm {wave,\perp}} $ illustrates the details of the field-particle interaction process responsible for the energy conversion from waves to ions. The positive correlation in phase between the bulk velocity vector of $\delta f_{\mathrm i}(\mathbf V)$ and $\delta \mathbf{B}_{\mathrm {wave,\perp}}$ in Figure 5 and Figure 6 is consistent with the anti-parallel propagation of ion cyclotron waves.

\begin{figure}
\centerline{\includegraphics[width=20cm, clip=]{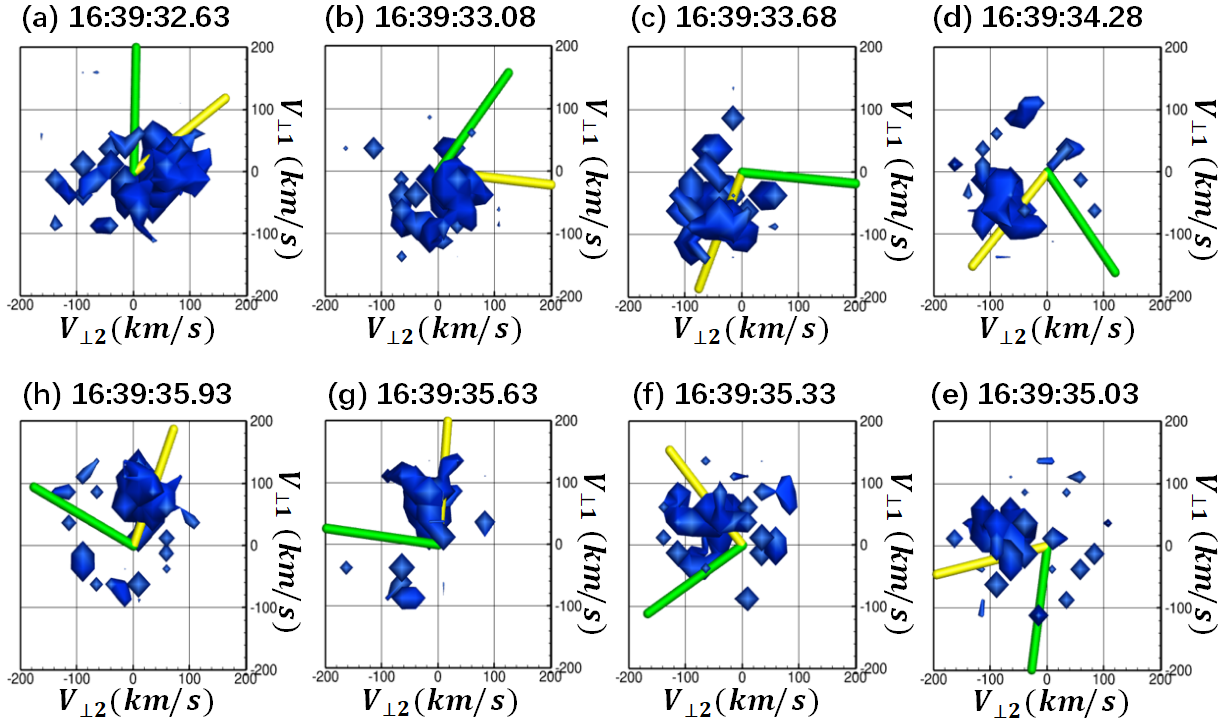}}
\caption{3-D contour surfaces of the fluctuating ion velocity distribution function ($\delta f_{\mathrm i}(\mathbf V)$) at different phases during one wave period. The contour levels of the  $\delta f_{\mathrm i}(\mathbf V)$ are selected as $5\times 10^{-23} (cm^{-6}s^3)$. The background magnetic field is in the out-of-plane direction. The wave electric field vectors and the magnetic field vectors are marked by the green and yellow sticks, respectively.
\label{fig:fig5}}
\end{figure}

\begin{figure}
\centerline{\includegraphics[width=20cm, clip=]{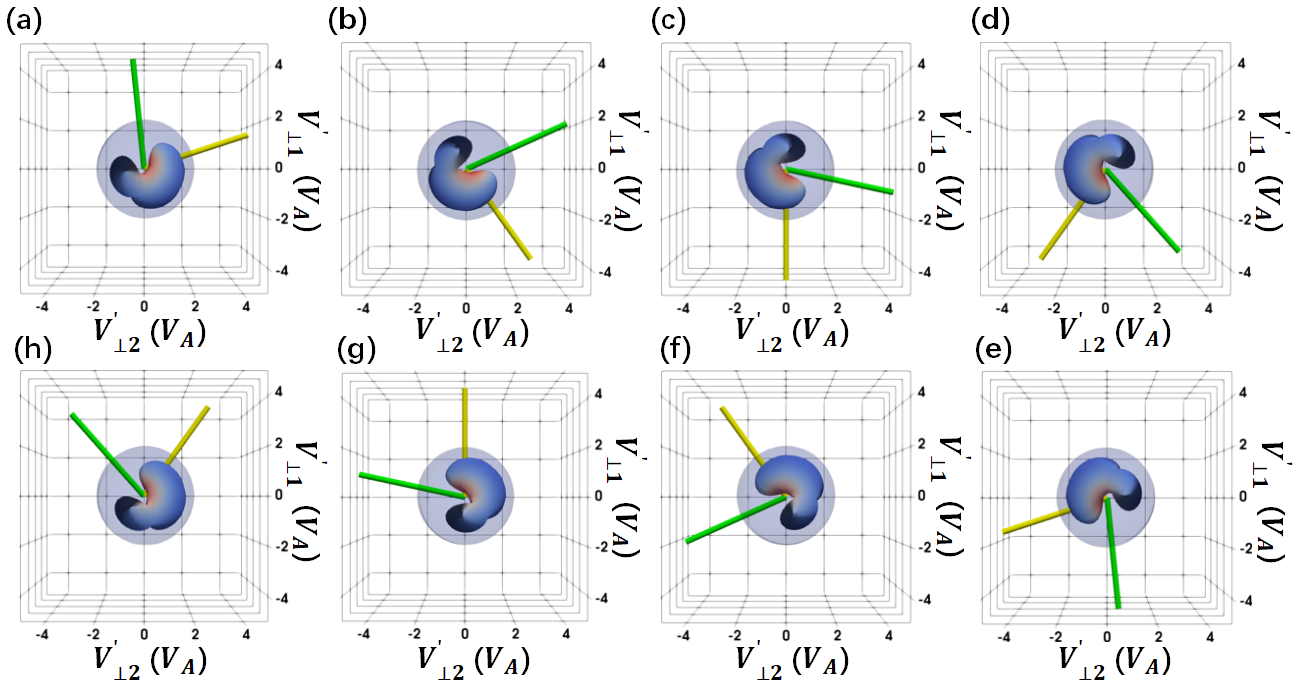}}
\caption{3-D contour surfaces of the fluctuating (opaque, $\delta f_{\mathrm i}(\mathbf V)$) and total (transparent, $f_{\mathrm i}(\mathbf V)$) ion phase space density at different phases during one wave period, based on the PKUES solver. The contour levels of the  $\delta f_{\mathrm i}(\mathbf V)$ and $f_{\mathrm i}(\mathbf V)$ are selected as $5\times 10^{-23} (cm^{-6}s^3)$ and $5\times 10^{-24} (cm^{-6}s^3)$, respectively. The background magnetic field is in the out-of-plane direction. The green and yellow sticks mark the electric and magnetic field vectors, respectively.
\label{fig:fig6}}
\end{figure}

\section{Summary and Discussion} \label{sec:result}
Using MMS’s measurements of particles and fields, we present the correlation between the fluctuating ion velocity distribution function ($\delta f_{\mathrm i}(\mathbf V))$ and wave electric field vectors $\delta \mathbf{E'}_{\mathrm {wave,\perp}}$, which is the essence of cyclotron resonance. The absolute relative phase angle defined as the azimuthal angle difference between the maximum of $\delta f_{\mathrm i}(=\delta f_{\mathrm {i,max}})$ and $\delta \mathbf{E'}_{\mathrm {wave,\perp}}$, |$\zeta$|=$|\phi(\delta f_{\mathrm {i,max}})-\phi(\delta \mathbf{E'}_{\mathrm {wave,\perp}})|$,  is less than $90\degree$, suggesting the energy conversion from wave fields to particles. Furthermore, the integrated work done by the electromagnetic field on ions is positive, indicating that ions are mainly energized by the perpendicular component of the electric field via cyclotron resonance. Therefore, our combined analysis of MMS observations and plasma wave theory provides direct and comprehensive evidence for ICW damping in space plasmas.

Since this work focuses on the kinetic energy conversion in the magnetosphere, the direct finding of ICW damping and thus the energy conversion from wave fields to particles is an important step towards the understanding of energy redistribution through field-particle interaction in collisionless plasma. Based on the fact that field-particle interaction exists widely in the heliosphere, the result of this work is of scientific significance, because it provides an observational basis supported by theoretical considerations as well as a physical scenario for the ICW damping and energy conversion in collisionless plasmas. This work also points out that even advanced plasma detectors like FPI onboard MMS need to be further improved to meet the needs of accurate measurement of ion velocity distribution in sparse plasmas like the magnetosphere. In this work, the most limiting factor is the geometric factor of the instrument.

\section{Acknowledgements}
The authors are grateful to the teams of the MMS spacecraft for providing the data. We also thank the team of 3DView, which is maintained mainly by IRAP/CNES. The work at Sun Yat-sen University is supported by NSFC through grant 2030201. The work at Peking University is supported by NSFC (No. 41874200 and 42174194) and National Key R\&D Program of China (No. 2021YFA0718600). D.V. from UCL is supported by the STFC Ernest Rutherford Fellowship ST/P003826/1 and STFC Consolidated Grant ST/S000240/1. This work is also supported by CNSA under contracts No. D020301 and D020302 and supported by the Key Research Program of the Institute of Geology \& Geophysics, CAS, Grant IGGCAS‐ 201904. XYZ is the co-first author of this work.





\end{document}